\input harvmac

\def\frak#1#2{{\textstyle{{#1}\over{#2}}}}
\def\frakk#1#2{{{#1}\over{#2}}}
\def\pa{\partial}
\def\Ycal{{\cal Y}}
\def\semi{;\hfil\break}

\def\Ytilde{\tilde Y}
\def\NSVZ{{\rm NSVZ}}
\def\DRED{{\rm DRED}}

\def\npb{{Nucl.\ Phys.\ }{\bf B}}

\def\prd{{Phys.\ Rev.\ }{\bf D}}
\def\prl{Phys.\ Rev.\ Lett.\ }
\def\plb{{Phys.\ Lett.\ }{\bf B}}

\def\thbar{\bar\theta}
\def\mbar{{\bar m}}

\def\bxhat{\hat\beta_{\xi}}
\def\lf{16\pi^2}
\def\llf{(16\pi^2)^2}
\def\lllf{(16\pi^2)^3}
{\nopagenumbers
\line{\hfil LTH 468}
\line{\hfil hep-ph/9911491}
\vskip .5in    
\centerline{\titlefont Renormalisation of the Fayet-Iliopoulos $D$-term}
\vskip 1in
\centerline{\bf I.~Jack, D.R.T.~Jones}
\medskip
\centerline{\it Dept. of Mathematical Sciences,
University of Liverpool, Liverpool L69 3BX, UK}
\vskip .3in

We consider the renormalisation of the  Fayet-Iliopoulos $D$-term in a
softly-broken  Abelian supersymmetric theory. We show that there exists
(at least through three loops)  a renormalisation group invariant 
trajectory for the coefficient of the $D$-term,  corresponding to the
conformal anomaly solution for the soft masses and couplings. 
 
\Date{December 1999}

\newsec{Introduction}

In Abelian gauge theories with $N=1$ supersymmetry 
there exists a possible 
invariant that is not allowed in the non-Abelian case: the 
Fayet-Iliopoulos $D$-term,
\eqn\dta{L =
\xi\int V (x,\theta, \thbar)\,d^4\theta  = \xi D(x).} 
In this paper we discuss the 
renormalisation of $\xi$ in the presence of 
the standard soft supersymmetry-breaking terms 
\eqn\Aaf{
L_{\rm SB}=(m^2)^j_i\phi^{i}\phi_j+
\left(\frak{1}{6}h^{ijk}\phi_i\phi_j\phi_k+\frak{1}{2}b^{ij}\phi_i\phi_j
+ \frak{1}{2}M\lambda\lambda+{\rm h.c.}\right)}

Let us begin by reviewing the position when there is no
supersymmetry-breaking,  i.e. for $L_{\rm SB}=0$.  Many years ago,
Fischler et al\ref\fisch{W. Fischler et al, \prl 47 (1981) 757}  proved
an important  result concerning  the renormalisation of the $D$-term
(see also  Ref.~\ref\shwg{M.A. Shifman and A.I. Vainshtein \npb 277
(1986) 456\semi S.~Weinberg, \prl 80 (1998) 3702}).   Since it is a
$\int\,d^4\theta$-type term, one may expect that the $D$-term  will
undergo renormalisation in general. Moreover, by simple power-counting 
it is easy to show that the said renormalisation is in  general
quadratically divergent. Evidently this poses a naturalness  problem
since (if present) it would introduce the cut-off mass scale  into the
scalar potential. At the one loop level it is easy  to show that the
simple condition $\Tr \Ycal = 0$ (where $\Ycal$ is  the $U_1$
hypercharge and the trace is taken over the chiral supermultiplets) 
removes the divergence. Remarkably, although one may of course  easily
draw individual diagrams proportional (for example)  to $\Tr \Ycal^5,
\Ycal^7 \cdots$ etc.,  this condition suffices to all orders.

In the presence of supersymmetry breaking, however, it is clear that $\xi$ will 
suffer {\it logarithmic\/} divergences. If calculations are done  in the
component formalism with $D$ eliminated by means of its equation of 
motion, then these divergences are manifested via contributions to  the
$\beta$-function for $m^2$. It is in this
manner that the results for the soft $\beta$-functions were given 
in, for example,  Ref.~\ref\jj{I.~Jack and D.R.T.~Jones, \plb 333 (1994)
372}.  Here we prefer to consider the renormalisation of $\xi$
separately; an  advantage of this is that it means that the exact
results for the soft $\beta$-functions presented in 
Refs.~\ref\jjf{I.~Jack and D.R.T.~Jones, \plb 415 (1997) 383}%
\nref\akk{L.V.~Avdeev, D.I.~Kazakov and I.N.~Kondrashuk, 
\npb510 (1998) 289}%
--\ref\jjp{I.~Jack, D.R.T.~Jones and A.~Pickering,  \plb 426 (1998) 73;
{\it ibid\/} 432 (1998) 114; {\it ibid\/} 435 (1998) 61}\ 
(see also Ref.~\ref\ahetal{ J. Hisano and M. Shifman, 
\prd 56 (1997) 5475\semi N.~Arkani-Hamed et al,
\prd58 (1998) 115005})  
apply
without change to the Abelian case.
The result for $\beta_{\xi}$ is as follows:
\eqn\exacta{\beta_{\xi} = \frakk{\beta_g}{g}\xi + \bxhat}
where $\bxhat$ is determined by $V$-tadpole (or in components 
$D$-tadpole) graphs, and is independent of $\xi$. In the 
supersymmetric case, we have $\bxhat = 0$, whereupon Eq.~\exacta\ is equivalent 
to the statement that the $D$-term, Eq.~\dta, is unrenormalised. 
In the presence of Eq.~\Aaf, however, $\bxhat$ depends on 
$m^2$, $h$ and $M$ (it is easy to see that it cannot depend on $b$.)
It is interesting that the dependence on  $h$ and $M$ arises 
first at the three loop level.  

Although in this paper we restrict ourselves to the Abelian case, 
it is evident that a $D$-term can occur with a direct product gauge group
($G_1\otimes G_2\cdots$) if there is an  Abelian factor: 
as is the case for the  MSSM. In the MSSM context one may 
treat $\xi$ as a free parameter at the weak scale\ref\gouv{ A. de Gouv\^ea,
A. Friedland and H. Murayama, \prd 59 (1999) 095008}, 
in which case there is no need to know $\bxhat$. However, if 
we know $\xi$ at gauge unification, then we need $\bxhat$ to predict 
$\xi$ at low energies.  Now in the $D$-uneliminated case it is possible 
to express all the $\beta$-functions associated with the soft 
supersymmetry-breaking terms given in Eq.~\Aaf\ in terms of the gauge 
$\beta$-function $\beta_g$, the chiral supermultiplet anomalous 
dimension $\gamma$ and a certain function $X$ which appears only 
in $\beta_{m^2}$; moreover in a special renormalisation 
scheme (the NSVZ scheme), $\beta_g$ can also be 
expressed in terms of $\gamma$. 
It is clearly of interest to ask whether an analogous exact expression
exists for $\beta_{\xi}$. Moreover, there exists an exact solution 
to the soft RG equations for $m^2$, $M$ and $h$ corresponding to 
the case when all the supersymmetry-breaking arises from the 
conformal anomaly\ref\con{L. Randall and R. Sundrum, hep-th/9810155\semi
G.F. Giudice, M.A. Luty, H. Murayama and  R. Rattazzi,
JHEP 9812 (1998) 27\semi
A. Pomarol and  R. Rattazzi, JHEP 9905 (1999) 013\semi
T. Gherghetta, G.F. Giudice, J.D. Wells, hep-ph/9904378\semi
Z. Chacko, M.A. Luty, I. Maksymyk and E. Ponton, hep-ph/9905390\semi
E. Katz, Y. Shadmi and Y. Shirman, JHEP 9908 (1999) 015\semi
I. Jack and D.R.T.~Jones, \plb 465 (1999) 148\semi
J.A.~Bagger, T. Moroi and E. Poppitz, hep-th/9911029}
and it is also interesting to ask whether this solution can be 
extended to the non-zero $\xi$ case.

The key to the derivation of the exact results for the soft $\beta$-functions
is the spurion formalism. The obstacle to deriving an analogous result for 
$\beta_{\xi}$ is the fact that individual superspace diagrams are (as already 
mentioned) quadratically divergent. This means that if, for example, 
we represent a $h^{ijk}$ vertex in superspace by $h^{ijk}\theta^2$, then 
we cannot simply factor the $\theta^2$ out, because it can be 
``hit'' by a superspace $D$-derivative. The simple relationship between 
a graph with a $h^{ijk}$ and the corresponding one with 
a supersymmetric Yukawa vertex which holds for the soft breaking
$\beta$-functions  is thereby lost. We are therefore unable 
to construct an exact formula for $\beta_{\xi}$; we do, however, present 
a solution for $\xi$ related to the conformal anomaly solution, but which 
must be constructed order by order in perturbation theory.     

\newsec{The $\beta$-function for $\xi$}

In this section we derive Eq.~\exacta, and  show how the contributions
to $\bxhat$ proportional to $m^2$ can be related  in a simple way to
$\beta_g$. 

We take an Abelian  $N=1$ supersymmetric gauge theory with
superpotential
\eqn\eqf{W(\Phi)=\frak{1}{6}Y^{ijk}\Phi_i\Phi_j\Phi_k+   
\frak{1}{2}\mu^{ij}\Phi_i\Phi_j,}
and at one loop we have 
\eqna\beone$$\eqalignno{\lf\beta_g^{(1)} &= g^3Q = 
g^3\Tr\left[\Ycal^2\right], &\beone a\cr
\lf\gamma^{(1)i}{}_j &= P^i{}_j 
=\frak{1}{2}Y^{ikl}Y_{jkl}-2g^2(\Ycal^2)^i{}_j. 
&\beone b\cr}$$

Let us consider the $D$-term renormalisation. 
We define renormalised and bare quantities in the usual way:
\eqn\appa{ \xi_B\int V_B\,d^4\theta  = 
\mu^{-\frakk{\epsilon}{2}}\xi Z_{\xi}Z_V^{\frakk{1}{2}}\int V \,d^4\theta 
}
where the $\mu$ factor establishes the canonical 
dimension for $\xi$ in $d = 4 - \epsilon$ dimensions. 
Then writing 
\eqn\appb{\xi Z_{\xi}Z_V^{\frakk{1}{2}} = \xi + \sum \frakk{a_n}{\epsilon^n}}
and using
\eqn\appc{\eqalign{\xi_B &= \mu^{-\frakk{\epsilon}{2}}\xi Z_{\xi}\cr
g_B &= \mu^{\frakk{\epsilon}{2}}g Z_V^{-\frakk{1}{2}} \cr}}
it is straightforward to show, using 
\eqn\appd{\mu\frakk{\pa g}{\pa\mu} = -\frakk{\epsilon}{2}g + \beta_g}
that 
\eqn\appe{\mu\frakk{\pa \xi}{\pa\mu} = \frakk{\epsilon}{2}\xi + \beta_{\xi}}
where 
\eqn\appf{\beta_{\xi} = \frakk{\beta_g}{g}\xi + \bxhat,}
and  where 
\eqn\appg{\bxhat = \sum_L L a_1^L.} 
Here $a_1^L$ is the contribution to $a_1$ from diagrams with $L$ loops. 
On dimensional grounds, 
\eqn\apph{\bxhat = m^2 A_1(g, Y, Y^*) + hh^* A_2 (g, Y, Y^*)
+MM^* A_3(g, Y, Y^*) + (Mh^* + M^*h)A_4 (g, Y, Y^*),}
where we have suppressed $(i,j\cdots)$ indices for simplicity.

By considering the relationship between the original theory 
and the one obtained by elimination of the $D$-field, we can 
prove a remarkably simple result for $A_1$ above. 
The relevant part of the supersymmetric Lagrangian is as follows:
\eqn\lsusy{
L = \frak{1}{2}D^2 +  \xi D + gD\phi^*\Ycal\phi - \phi^* m^2\phi+\cdots .}
After eliminating $D$ this becomes
\eqn\lsusya{
L = -\phi^* \mbar^2\phi - \frak{1}{2}g^2(\phi^*\Ycal\phi)^2,}
where 
\eqn\lsusyb{
 \mbar^2 = m^2 + g\xi \Ycal.}
RG invariance of this result gives  (using Eq.~\appf)
\eqn\lsusyc{
\beta_{\mbar^2}(\mbar^2, \cdots) = \beta_{m^2}(m^2, \cdots) 
+ 2\beta_g\xi \Ycal + g \Ycal \bxhat(m^2,\cdots).}
Now $\beta_{m^2}$ is calculated in the uneliminated Lagrangian 
and hence does not contain the ``$D$-tadpole'' contributions. It is, in fact, 
given precisely by the previously derived formula:
\eqn\Ajy{
(\beta_{m^2})^i{}_j (m^2, \cdots) =\left[ \Delta + 
X\frakk{\pa}{\pa g}\right]
\gamma^i{}_j.}
where
\eqn\Ajz{
\Delta = 2{\cal O}{\cal O}^* +2MM^{*} g^2{\pa
\over{\pa g^2}} +\Ytilde_{lmn}{\pa\over{\pa Y_{lmn}}}
+\Ytilde^{lmn}{\pa\over{\pa Y^{lmn}}},}
\eqn\Ajb{
{\cal O}=\left(Mg^2{\pa\over{\pa g^2}}-h^{lmn}{\pa
\over{\pa Y^{lmn}}}\right),}
\eqn\Ajd{
\Ytilde^{ijk}=(m^2)^i{}_lY^{ljk}+(m^2)^j{}_lY^{ilk}+(m^2)^k{}_lY^{ijl}}
and (in the NSVZ scheme) 
\eqn\exX{
\lf X^{\NSVZ}=-2g^3 \Tr \left[m^2\Ycal^2\right].} 
Now $\beta_{\mbar^2}$ is given by 
\eqn\appi{
\beta_{\mbar^2} = \beta_{m^2} (\mbar^2,\cdots) 
+ g\Ycal \bxhat(\mbar^2, \cdots).}
In other words, if we calculate in the $D$-eliminated formalism, we obtain
both ``normal'' contributions (the ones that would be single-particle 
irreducible in the $D$-uneliminated case) and the $D$-tadpole contributions, 
which appear in   $\beta_{\xi}$ in the $D$-uneliminated case.

The key now is the result that 
\eqn\appj{
\beta_{m^2} (\mbar^2,\cdots) = \beta_{m^2} (m^2,\cdots)}
This follows simply by substituting for $\mbar^2$ from Eq.~\lsusyb\ and 
then using the facts that 
\eqn\appk{
(\Ycal)^i{}_lY^{ljk}+(\Ycal)^j{}_lY^{ilk}+(\Ycal)^k{}_lY^{ijl} = 0}
by gauge invariance, and 
\eqn\appl{\Tr (\Ycal^3) = 0}
for anomaly cancellation.\foot{A remark on scheme dependence. 
The result for $X$, 
Eq.~\exX, applies in the NSVZ scheme, which is one of a class of schemes 
(which include the standard perturbative method, DRED),
related by redefinitions of $g$ and $M$, the ramifications of which 
are described in Refs.~\jjp. Now $X$ transforms non-trivially 
under these redefinitions, but it can be shown using Eqs.~\appk, 
\appl\ that $X$ is unchanged by the replacement 
$m^2\to \mbar^2$ in any member of this class of schemes; 
consequently Eq.~\appj\ always applies.}
We then find immediately that:
\eqn\appx{
\bxhat(\mbar^2,\cdots) = 2\frakk{\beta_g}{g}\xi + \bxhat(m^2,\cdots)}
whence 
\eqn\appm{ \Tr(\Ycal A_1) = 2\frakk{\beta_g}{g^2}.}
So if we take the $D$-tadpole contributions to $\beta_{\xi}$, then 
the terms proportional to $m^2$ will reduce to $2 \beta_g/g$ if we replace 
$m^2$ by $g\Ycal$. This result is, in fact, clear from a diagrammatic 
point of view, since the aforesaid replacement converts the diagrams 
into $D$ self-energy graphs, and hence indeed gives rise to $\beta_g$.

\newsec{The three loop results}

Through two loops we have that
\eqn\exactb{\lf\bxhat = 2g\Tr\left[\Ycal m^2\right]
-4g\Tr\left[\Ycal m^2 \gamma^{(1)}
\right]+\cdots}
so we see that in fact only $A_1$ is non-zero through this order. 
Moreover, since
\eqn\thrb{
\lf\beta_g = g^3\Tr\left[\Ycal^2\right] 
-2g^3\Tr\left[\Ycal^2\gamma^{(1)}\right]+\cdots}
we see that Eq.~\exactb\ is indeed consistent with Eq.~\appm.

We have calculated  
several distinct gauge invariant contributions to 
$\bxhat^{(3)\DRED}$, namely the sets 
of terms that are $O(gY^4 m^2)$, $O(g Y^2 h^2)$, 
$O(g^3Y^2 m^2)$ and $O(g^3 h^2)$. 
We find that:
\eqn\exactc{\eqalign{\lllf\frakk{\bxhat^{(3)\DRED}}{g} &= 
7(Y^2)^i{}_j Y^{jkl}Y_{ikm}(m^2\Ycal)^m{}_l 
+4(Y^2)^i{}_j Y^{jkl}Y_{imn}(m^2)^m{}_k\Ycal^n{}_l\cr& 
- \frak{3}{2}\Tr\left[ Y^2 Y^2 m^2 \Ycal\right]
- \frak{5}{2}Y^{ikl}Y_{imn}h_{jkl}h^{pmn}\Ycal^p{}_j 
- 2\Tr\left[ Y^2 h^2 \Ycal\right]\cr&
 +(10-24\zeta(3))g^2\Tr\left[ Y^2 m^2\Ycal^3\right]
-12g^2Y^{ikl}Y_{imn}(m^2\Ycal)^m{}_k (\Ycal^2)^n{}_l\cr& 
+ (8-24\zeta(3))g^2 \left[2Y^{ikl}Y_{imn}(m^2)^m{}_k(\Ycal^3)^n{}_l
+h^{ikl}h_{jkl}(\Ycal^3)^j{}_i\right] +\cdots\cr}}
where 
$(Y^2)^i{}_j =  Y^{ikl}Y_{jkl}, (h^2)^i{}_j =  h^{ikl}h_{jkl}.$
Then replacing 
$m^2$ by $g\Ycal$, we obtain
\eqn\appmn{g\lllf\Tr(\Ycal A_1^{(3)}) =  (6X_1 + 12X_3+ 2X_4)+\cdots,}
where 
\eqn\inv{\eqalign{X_1&= g^2Y^{klm}P^n{}_l(\Ycal^2)^p{}_mY_{knp},
\cr
X_3 &= g^4\Tr \left[P\Ycal^4\right],\cr
X_4 &= g^2\Tr\left[P^2 \Ycal^2\right],\cr}} 
in precise agreement with the result for $\beta_g^{(3)}$, given in 
\ref\jjna{I.~Jack, D.R.T.~Jones and C.G.~North, 
\plb386 (1996) 138}, which for an Abelian theory is 
\eqn\bthree{\lllf\beta_g^{(3)\DRED}= g\left\{3X_1+6X_3+X_4-6g^6Q\Tr
\left[\Ycal^4\right]\right\}.}
We have not calculated the $O(g^5 m^2)$ contributions to 
$\bxhat^{(3)}$, 
which will produce the $O(g^7)$ terms in Eq.~\bthree. 
 
\newsec{The conformal anomaly trajectory}

The following set of equations provide an exact solution to the 
renormalisation group equations for $M, h, b$ and $m^2$:
\eqna\result$$\eqalignno{M &= M_0{\beta_g\over g}, &\result a\cr
h^{ijk}&=-M_0\beta_Y^{ijk},&\result b\cr
b^{ij}&=-M_0\beta_{\mu}^{ij}, &\result c\cr
(m^2)^i{}_j &= \frak{1}{2}|M_0|^2\mu\frakk{d\gamma^i{}_j}{d\mu}.
&\result d\cr}$$
Moreover, these solutions indeed hold if the only source of 
supersymmetry breaking is the conformal anomaly, when $M_0$ is 
in fact the gravitino mass.

This set of soft breakings has caused considerable interest; 
but there are clear difficulties for the MSSM, since  it is
easy to see that sleptons are predicted to have  negative
$\hbox{(mass)}^2$.  Most studies of this scenario have resolved this
dilemma by adding a constant $m_0^2$, presuming another source of
supersymmetry breaking. A non-zero $\xi$ alone is not an alternative,
unfortunately, as is  easily seen from Eq.~\lsusyb; the two selectrons,
for example, have  oppositely-signed 
hypercharge so one of them at least remains
with  negative $\hbox{(mass)}^2$.  

It is immediately obvious that, given Eq.~\result{}, there is a RG invariant 
solution for $\xi$ through two loops (for $\bxhat$) given by: 
\eqn\exactxi{\lf\xi = g|M_0|^2\Tr\left[\Ycal (\gamma-\gamma^2)\right],} 
since differentiating with respect to $\mu$ and using 
Eq.~\result{d}\  leads at once to Eqs.~\appf, \exactb. 
Interestingly, however, this result for $\xi$ {\it vanishes\/} 
at leading  and next-to-leading order, since one easily demonstrates that 
\eqn\simpa{\Tr\left[\Ycal \gamma^{(1)}\right] = 0}
and  
\eqn\simpb{\Tr\left[\Ycal \gamma^{(2)}\right] =    
\Tr\left[\Ycal (\gamma^{(1)})^2\right],}
using the result for $\gamma^{(2)}$, which is 
\eqn\gamtwo{\llf\gamma^{(2)i}{}_j =\left[-Y_{jmn}Y^{mpi}-2g^2(\Ycal^2)^p{}_j
\delta^i{}_n\right]P^n{}_p+2g^4(\Ycal^2)^i{}_jQ.}
It is interesting to ask 
whether the trajectory can be extended beyond two loops, and whether  it
in fact continues to vanish order by order. We have shown  that there is
indeed a generalisation of Eq.~\exactxi\ to at least three  
loops (for $\bxhat$), and
that at this order the result for $\xi$ is non-zero.

Our result is as follows: 
\eqn\exactxib{\eqalign{\frakk{\lf\xi^{\DRED}}{g|M_0|^2} &=
\Tr\left[\Ycal (\gamma-\gamma^2)\right]
-24\zeta(3)g^2(\lf)^{-3}\left[\Tr\left[\Ycal^3 P^2\right]
+ (\Ycal^3)^i{}_jY^{jkl}Y_{ikm}P^m{}_l\right]\cr
&+2(\lf)^{-3}\left[(\Ycal)^i{}_j Y^{jkl}Y_{imn}P^m{}_k P^n{}_l
-2\Tr\left[\Ycal P^3\right]-4g^2\Tr\left[\Ycal^3 P^2\right]\right]
+\cdots.\cr}}
Note that our partial $\bxhat$ expression determines only the  $O(Y^6)$
and $(Y^4g^2)$ terms on the RHS of Eq.~\exactxib;  it is interesting
that they can be written in terms of invariants involving  $P$. After
this is done we have chosen to ``promote'' remaining  $Y^2$ factors to
$2P$; the difference thereby introduced depends  only on terms we have
not calculated.    It is easy to verify that (for $Y^8$ and $Y^6g^2$
terms)  the result of  taking $\mu\frakk{\pa}{\pa\mu}$ of this
expression is  identical to that obtained by substituting
Eq.~\result{b,d}\ in Eqs.~\exactb, \exactc. This is a non-trivial result
in that  the number of candidate terms  for inclusion in Eq.~\exactxib\
is  considerably  less than the number of distinct terms which arise
when Eq.~\exactb, \exactc\  are placed on the RG trajectory. We
therefore conjecture that the  trajectory holds for the full
$\bxhat^{(3)}$ calculation, and extends  to all orders.  

Using the result\foot{
Notice that 
$\Tr\left[ \Ycal\gamma^{(3)\DRED}\right] = 
\Tr\left[ \Ycal\gamma^{(3)\NSVZ}\right]$ in the Abelian case, 
by virtue of Eq.~\appl.} for 
$\gamma^{(3)}$ from Ref.~\ref\jjnb{
I.~Jack, D.R.T.~Jones and C.G.~North, \npb486 (1997) 479},
we find that 
\eqn\conject{\lllf\Tr\left[\Ycal (\gamma-\gamma^2)\right] =
I_1 + 12\zeta(3)I_2 + O(Y^8, Y^6g^2, \cdots g^8),}where
\eqn\conjectb{\eqalign{I_1 &= \Tr\left[\Ycal P^3\right]
-\frak{1}{2}(\Ycal)^i{}_j Y^{jkl}Y_{imn}P^m{}_k P^n{}_l
+ 2g^2 \Tr\left[\Ycal^3 P^2\right]
-2g^4Q\Tr\left[\Ycal^3 P\right]\cr
I_2&= g^2(\Ycal^3)^i{}_jY^{jkl}Y_{ikm}P^m{}_l+
g^2\Tr\left[\Ycal^3 P^2\right] + 2g^4\Tr\left[\Ycal^5 P\right].\cr}}
Comparing Eq.~\exactxib\ with Eq.~\conjectb\ leads us to the conjecture 
that on the RG trajectory $\xi^{\DRED}$ is given at leading order  
by the expression 
\eqn\conjectc{\frakk{\xi^{\DRED}}{g|M_0|^2} = (\lf)^{-4}\left(-3I_1 
-12\zeta(3)I_2 + \nu_1Qg^6\Tr\left[\Ycal^5\right] 
+ \nu_2Qg^4\Tr\left[P\Ycal^3\right]\right)}
where $\nu_1, \nu_2$ are undetermined. Note that other invariants which might
in principle have appeared in Eq.~\conjectc\ are in fact ruled out from 
consistency with $\bxhat^{(3)\DRED}$; and in fact our 
conjecture is equivalent to the following form for 
$\bxhat^{(3)\DRED}$:
\eqn\conjectd{\eqalign{
\lllf\frakk{\bxhat^{(3)\DRED}}{g}&=-6(\lf)^2\Tr\left[\Ycal m^2
\gamma^{(2)}\right]-4\Tr\left[WP\Ycal\right]- 
\frak52\Tr\left[HH^*\Ycal\right]\cr&
+2\Tr\left[P^2m^2\Ycal\right]
-24g^2\zeta(3)\Tr\left[W\Ycal^3\right]\cr&
+\left(12\zeta(3) - \frak{1}{2}\nu_2\right)g^2\Tr\left[M^*H\Ycal^3 
+ \hbox{c.c.}\right]\cr&
+6\left(-48\zeta(3)+ \nu_1\right)g^4MM^*\Tr\left[\Ycal^5\right]
+4\nu_2 g^2MM^*\Tr\left[P\Ycal^3\right]\cr
}}
where\jj 
\eqn\betam{\eqalign{
W^i{}_j =
(\frak{1}{2}Y^2m^2 +\frak{1}{2}m^2Y^2 +h^2)^i{}_j
+2Y^{ipq}Y_{jpr}(m^2)^r{}_q -8g^2MM^*(\Ycal^2)^i{}_j\cr}}
and
\eqn\newX{
H^i{}_j=h^{ikl}Y_{jkl}+4g^2M(\Ycal^2)^i{}_j.}
It is interesting to note that the particular value 
$\nu_1 = 48\zeta(3)$ makes $\bxhat^{(3)\DRED}$ vanish 
if $P^i{}_j=0$, $(m^2)^i{}_j=\frak13MM^*\delta^i{}_j$, and 
$h^{ijk}=-MY^{ijk}$. These relations arise 
(at leading order) in finite softly-broken theories (of course 
in the Abelian case considered here we cannot achieve a finite theory 
as $Q\neq 0$).    

It is natural to ask what the result for $\bxhat^{(3)}$ is in the NSVZ 
scheme, which is obtained (at the relevant order) by the 
redefinitions\jjf
\eqn\drnz{\eqalign{
\llf\delta g &= -\frak{1}{2}g^3\Tr\left[P\Ycal^2\right]\cr
\llf\delta M &= -Mg^2\left\{\Tr\left[P\Ycal^2\right]
-2g^2\Tr\left[(\Ycal^2)^2\right]\right\}
+\frak{1}{2}g^2 h^{ikl}Y_{jkl}(\Ycal^2)^j{}_i.\cr}} 
It is straightforward to show that in order to obtain the results
Eqs.~\appf\ and \appm\ in the NSVZ scheme,  
we must also redefine $\xi$ as follows:
\eqn\xirdfn{
\llf\delta \xi = -\frak{1}{2}g^2\Tr\left[P\Ycal^2\right]\xi
-g \Tr\left[m^2P\Ycal\right].}
The effect of this is to replace Eq.~\conjectc\ by 
\eqn\conjectcd{\frakk{\xi^{\NSVZ}}{g|M_0|^2} = (\lf)^{-4}\left(-4I_1
-12 \zeta(3)I_2+ \nu_1Qg^6\Tr\left[\Ycal^5\right] 
+ \nu_2Qg^4\Tr\left[P\Ycal^3\right]\right),}
and Eq.~\conjectd\ by 
\eqn\exactcd{\eqalign{\lllf\frakk{\bxhat^{(3)\NSVZ}}{g} 
&=-4(\lf)^2\Tr\left[\Ycal m^2 
\gamma^{(2)}\right]
-\frak{5}{2}\left(2\Tr\left[WP\Ycal\right]+
\Tr\left[HH^*\Ycal\right]\right)\cr
&-24g^2\zeta(3)\Tr\left[W\Ycal^3\right]
+\left(12\zeta(3) - \frak{1}{2}\nu_2\right)g^2\Tr\left[M^*H\Ycal^3 
+ \hbox{c.c.}\right]\cr&
+6\left(-48\zeta(3)+ \nu_1\right)g^4MM^*\Tr\left[\Ycal^5\right]
+4\nu_2 g^2MM^*\Tr\left[P\Ycal^3\right].\cr
}}
It is interesting to note that with the particular values
$\nu_1=48\zeta(3)$, $\nu_2= - 24\zeta(3)$, 
Eqs.~\exactb, \exactcd\ are consistent 
with the following expression  
\eqn\conjectdd{\eqalign{
\lf\frakk{\bxhat^{\NSVZ}}{g}
&=2\Tr\left[\Ycal m^2\right]-4\Tr\left[\Ycal m^2 \gamma\right]\cr& 
+\left(\Delta+X{\pa\over{\pa g}}\right)
\left(-\frak52\Tr[\gamma^2\Ycal]+24\zeta(3)g^2(\lf)^{-1}
\Tr[\gamma\Ycal^3]\right),}}
where $\Delta$ and $X$ are defined in Eqs.~\Ajy--\exX. 

We hope to test our conjectures Eq.~\conjectc\ and \conjectd\ 
by completing the 
calculation of $\bxhat^{(3)}$; we also plan to discuss scheme 
dependence in more detail,  and extend our results to a gauge group with
direct product  structure and  one or more abelian factors (such as the
MSSM).

\noindent{\bf Note added}

Since this paper was submitted we have calculated the contributions to 
$\bxhat^{(3)\DRED}$ of the form  $g^3\Tr\left[M^*H\Ycal^3  +
\hbox{c.c.}\right]$ and  $g^3MM^*\Tr\left[P\Ycal^3\right]$.  These are
{\it both\/} consistent with the result $\nu_2 = 0$.  Although this
means that the $\zeta(3)$ terms cannot, in fact, be written in the form 
suggested in Eq.~\conjectdd,  it does provide strong evidence in favour
of our  result Eq.~\conjectd. We feel that our  demonstration that $\xi$
has a RG trajectory related to the conformal  anomaly one is intriguing,
and worthy of further investigation.  

\bigskip\centerline{{\bf
Acknowledgements}}\nobreak
 
This work was supported in part by a Research Fellowship from the
Leverhulme Trust. We thank John Gracey for conversations. 

\listrefs
\end